\documentclass[11pt]{article}
\usepackage{array}
\usepackage{graphicx}
\usepackage{cite}
\usepackage{textpos}
\usepackage{bbold}
\usepackage{adjustbox}

\title{ {Gauge Symmetry in Shape Dynamics
\footnote{Invited contribution to C. N. Yang 100 birthday volume, to be published by World Scientific.}
}
\\
\author {Frank Wilczek  \\
\small\it Center for Theoretical Physics, MIT, Cambridge, MA 02139 USA; \\
\small\it T. D. Lee Institute and Wilczek Quantum Center, \\
\small\it Shanghai Jiao Tong University, Shanghai, China;\\
\small\it Arizona State University, Tempe, AZ, USA; \\
\small\it Stockholm University, Stockholm, Sweden }}

\begin{document}

\maketitle

\begin{textblock*}{5cm}(11cm,-8.2cm)
\fbox{\footnotesize MIT-CTP/5437}
\end{textblock*}

\begin{abstract}
C. N. Yang's ideas about local gauge symmetry and non-integrable phases have been enormously fertile sources of inspiration in fundamental physics and in the quantum theory of matter.  They also arise naturally in describing the dynamics of deformable bodies.  Here I  extend previous discussions of the gauge symmetry of deformable bodies in several directions, bringing in an arbitrary number of dimensions, general time-dependence, conservation laws and adiabatic residuals.  I briefly indicate other potential applications of the conceptual framework.  
\end{abstract}
\medskip

\bigskip


Every theoretical physicist of my generation is a student of C. N. Yang.  His insights, notably including the clusters of ideas around parity violation \cite{parity}  and local gauge symmetry \cite{yang-mills}, permeate modern fundamental physics.  Beyond that, his career as as a whole epitomizes the synergy of truth and beauty that is the central miracle of physics \cite{yang_selecta}.  In this, as in the equations that bear his name, he is the worthy modern heir of James Clerk Maxwell.

\section{Gauge Structure: Fundamental, Emergent, Productive}

The principle of local gauge symmetry is central to our present theories of fundamental interactions.  It has been a guide to discovery, not only through its constructive role in explaining the existence and properties of force-mediating particles -- photons, $W$ and $Z$ bosons, color gluons, and gravitons
 -- as avatars of curvature, but also through the powerful constraints it imposes on possible interactions among fundamental fields.   Yet I think it is fair to say that gauge symmetry appears in relativistic field theory as a rather mysterious and even elusive gift.  For example, we are led to the equations of quantum chromodynamics by imposing local $SU(3)$ color gauge symmetry, but the observables of the theory are supposed to be gauge singlets, as are the asymptotic states (confinement).   The underlying gauge symmetry thereby vanishes from view, leaving behind a smile drawn in evanescent jets.

In recent years gauge symmetry has been identified in several other physical contexts, wherein it is not an irreducible axiom but rather an emergent consequence of other principles.  These include the circle of ideas around Berry's phase in basic quantum mechanics \cite{berry, wz, sw1}, now developed to support many sophisticated applications including topological band theory \cite{bernevig}; the statistical gauge fields of anyon theory \cite{lm, fw, asw}, recently exhibited directly in the quantum Hall effect \cite {anyon_collider, manfra} and being prepared for use in quantum information processing \cite{pan, microsoft}; and in the low-energy theory of quantum spin ice \cite{spin_ice}, among others.  This naturally raises the question whether gauge symmetries presently regarded as fundamental might emerge from more primitive structures, too.   In any case, such examples bring the aforementioned mystery: Why gauge symmetry? down to earth.  

Identification of gauge structure in a new problem is productive, because it brings in powerful conceptual tools -- {\it e. g}., covariant derivative, curvature, holonomy, topology -- that have been sharpened by decades of development and application in physics.   Here, the synergy of beauty and truth is tangible.

In this note I will discuss another example of emergent gauge symmetry, in the dynamics of deformable bodies \cite{sw2, montgomery, littlejohn, cabrera}.   We would like to calculate how the internal motion of a body affects its external position and orientation.  But if the shape of a body changes, its change in position is not unambiguously defined, since different parts move by different amounts.  To overcome this problem, we associate with each shape a corresponding ``reference shape'' that has a definite position and orientation.  The position and orientation of the body relative to its reference shape is then well-defined, and we can calculate it unambiguously.  Of course, the introduction of reference shapes introduces an element of convention into this procedure, because we might have chosen them differently.  As we will see, this element of convention is precisely a gauge structure.  Moreover the associated gauge field is highly non-trivial, and it appears prominently in the dynamical equation.

\section{Dynamical Equation for Deformable Bodies}

\subsection{Referenced Angular Momentum}

We consider an assembly of point masses $m^j$ whose configuration $x^j$ can change with time.   
For each configuration we have a reference shape, as described immediately above.  We define the time-dependent rotation $R$ that relates the configuration of the assembly to the corresponding reference shape configuration according to
\begin{equation}\label{shape_to_space}
x^j ~=~ Rs^j
\end{equation}
Then we have a dynamical equation that relate change of angular momentum $L^\alpha$ to external torque $\tau^\alpha$ according to
\begin{equation}\label{torque_equation}
\tau^\alpha ~=~ \frac{dL^\alpha}{dt} ~=~ \epsilon_{\alpha\beta\gamma} \, \sum\limits_j \, m^j \, R_{\beta\gamma} s^j_\rho \ \frac{d^2}{dt^2} \, (R_{\rho\sigma}s^j_\sigma)
\end{equation}

Our goal is to work Eqn.\,(\ref{torque_equation}) into a form that contains fewer variables and brings out its gauge symmetry.  Thus, in a broad sense, we will construct an effective theory of rotational motion that displays an emergent gauge symmetry.  We will first do this in a way that is valid in any number of dimensions (see {\it e. g.}, \cite{higher_D}) and emphasizes conceptual structure.  Then we will apply some algebraic tricks special to three dimensions, to make contact with more conventional presentations.  In the Appendix, as an exercise and sanity check, we get to the same dynamical equation by brute force algebra.  

\subsection{Inertia Tensor and  Angular Motion}

The primary object of interest, written now in a form valid in any number of dimensions, is the angular momentum
\begin{equation}
L_{\alpha \beta} ~\equiv~ \sum\limits_j \ m^j \, (\, x^j_\alpha  \, {\dot x}^j_\beta -  x^j_\beta  \, {\dot x}^j_\alpha )
\end{equation}
Bringing in the reference shapes using Eqn.\,(\ref{shape_to_space}), we are led define two shape-based contributions to $L$ according to
\begin{eqnarray}\label{shape_angular_momentum}
R^{-1}_{\ \alpha \rho} R^{-1}_{\ \beta \sigma} \ L_{\rho \sigma} ~&=&~ \sum\limits_j \, m^j \, (s^j_{\alpha} \,  (R^{-1} {\dot R} )_{\beta \gamma} \, s^j_\gamma \, - \,   s^j_{\beta} \,  (R^{-1} {\dot R} )_{\alpha \gamma} \, s^j_\gamma ) \nonumber \\
~&+&~  \sum\limits_j \  m^j \, (s^j_\alpha \, {\dot s}^j_\beta \, - \, s^j_\beta \, {\dot s}^j_\alpha ) \nonumber \\
~&\equiv&~ M^S_{\alpha \beta} \, + \, L^S_{\alpha \beta}
\end{eqnarray}
Here and throughout we use the superscript $^S$ to indicate objects defined using shape variables.  These allow us to extend the idea of a body-fixed frame to deformable bodies.  

Now let us analyze and simplify $M^S$.  Bringing in the angular velocity
\begin{equation}
R^{-1} \dot R ~\equiv~ \omega
\end{equation}
we have
\begin{equation}\label{momentum_velocity}
M^S_{\alpha \beta} ~=~ \sum\limits_j \, m^j \, ( s^j_{\alpha} \,  \omega_{\beta \gamma} \, s^j_\gamma \, - \, s^j_{\beta} \,  \omega_{\alpha \gamma} \, s^j_\gamma ) ~=~ I^S_{\alpha \beta; \gamma \eta} \, \omega_{\gamma \eta}
\end{equation}
where
\begin{equation}
 I^S_{\alpha \beta; \gamma \eta}  ~\equiv~  \frac{1}{2} \bigl( {\tilde I}^S_{\alpha \gamma} \delta_{\beta \eta} - {\tilde I}^S_{\alpha \eta} \delta_{\beta \gamma} - {\tilde I}^S_{\beta \gamma} \delta_{\alpha \eta} + {\tilde I}^S_{\beta \eta} \delta_{\alpha \gamma} \bigr) \end{equation} 
with
 \begin{equation}
{\tilde I}^S_{\alpha \beta} ~\equiv~  \sum\limits_j \, m^j \, s^j_\alpha \, s^j_\beta 
 \end{equation} 
 defines the inertia tensor.
 
 We can assign simple labels $A$ the antisymmetric pairings $(\alpha, \beta)$.  For $D=3$ these labels take three values, for $D=4$ six values, and so forth. In terms of those variables $I_{A;B}$ is a symmetric matrix, $\omega_B$ is a vector, and Eqn.\,(\ref{momentum_velocity}) becomes
\begin{equation}
M^S_A ~=~ I^S_{A;B} \omega_B
\end{equation}
or simply 
\begin{equation}
M^S~=~ I^S\, \omega
\end{equation}

For our purposes we also need to know that $I^S$, thus defined, is generally invertible.  To see that, consider the expectation value of $I^S$ relative to a ``vector'' $\zeta$.  We find
\begin{equation}\label{positivity}
\zeta_{\alpha \beta} \, I^S_{\alpha \beta; \gamma \eta} \, \zeta_{\gamma \delta} ~=~ 2 \sum\limits_j \, m^j  (\zeta_{\rho \sigma} s^j_\sigma) (\zeta_{\rho \tau} s^j_\tau)
\end{equation}
This is a sum of non-negative terms, each of which only vanishes when $s^j$ lies in the subspace $\zeta_{\rho \sigma} s^j_\sigma = 0$.  For assemblies of $s^j$ in general position the expectation value is positive, and in that case $I_{a;b}$ is a symmetric positive matrix.  This implies that it is not only invertible, but even diagonalizable with positive entries.   (Note however that the diagonalization process involves rotations in the ``index space''.   For $D > 3$ these might not correspond to rotations in the ambient $D$-dimensional space.)   Expressions related to Eqn.\,(\ref{positivity}) will occur in our discussion of conservation laws, below, where they represent quantities that are generically positive on physical grounds.

\subsection{Gauge Symmetry and Gauge Field}

$L^S$ has the form of an angular momentum in shape space.   To bring out its physical significance, however, we need to transcend the element of convention (gauge symmetry!) that enters into its construction.  We have the freedom to modify our choice of reference shapes according to
\begin{eqnarray}
s ~&\rightarrow&~ U(s)^{-1}\,  s \\
R ~&\rightarrow&~ R\, U(s)
\end{eqnarray}
This choice does not affect the angular momentum $L$ (with no superscript), but it does modify everything else in Eqn.\,(\ref{shape_angular_momentum}).  $R$ and $I^S$ change simply homogeneously, but $\omega, M^S$ and $L^S$ are tricker.  They bring in inhomogeneous terms, that must cancel between $M^S$ and $L^S$.  

Indeed, we have
\begin{equation}
\omega ~\rightarrow~ U^{-1}  \omega \, U \, -  \,  U^{-1} \, \dot U
\end{equation}
The required cancellation therefore implies that
\begin{equation}
\Omega dt ~ \equiv~ (I^{S})^{-1} \, L^S \, dt 
\end{equation}
transforms as an $so(3)$ gauge potential field on shape space \cite{guich}.  (Of course, its transformation law can also be verified directly.) 
To bring out the fact that $U$ is a function on shape space, we can introduce co-ordinates $\lambda^\kappa$ on that space, and write
\begin{equation}
s(t) ~\equiv~ s \bigl(\lambda^\kappa(t) \bigr) 
\end{equation}
\begin{equation}
\dot U ~=~ \frac{\partial U\bigl( s(\lambda) \bigr) }{\partial \lambda^\kappa} \, \frac{ d \lambda^\kappa}{dt}
\end{equation}
so that 
\begin{equation}
\omega dt ~\rightarrow~ U^{-1} \,  \omega dt \, U\, -   \, U^{-1} \frac{\partial U\bigl( s(\lambda) \bigr) }{\partial \lambda^\kappa} \, d \lambda^\kappa
\end{equation}
This gauge potential naturally appears as a one-form.  To make contact with conventional physics notation, we would write out components
\begin{equation}
\Omega \, dt~\equiv~ \Omega_\kappa \, d\lambda^\kappa
\end{equation}

This gauge structure brings in the $m^j$ as parameters, but it is dimensionless and purely a function of the shape space geometry.  It identifies the effective contribution to the angular velocity due to shape change, as opposed to collective motion.  The separation between those two contributions brings in an element of convention, i.e. a choice of gauge, but the total is free of convention, i.e. gauge covariant. 

Knowing that $\Omega\, dt$ is a gauge potential, we can use the techniques of gauge theory to bring out its gauge-invariant content.  We can calculate field strengths (curvature), Wilson loops (holonomy), and so forth.  In the present context, this allows us to 
identify unambiguous, quantitative dynamical consequences of shape deformation.

Using $\Omega$ we can write 
\begin{equation}
R^{-1}_{\alpha \mu} R^{-1}_{\beta \nu} L_{\mu \nu} ~=~ I^S_{\alpha \beta; \mu \nu}   ( \omega_{\mu \nu} + \Omega_{\mu \nu} ) 
\end{equation}
or more compactly
\begin{equation}
L ~=~ {\cal R} \, I^S (\omega + \Omega)
\end{equation}

\subsection{Dynamical Equation}

It is straightforward to pass from $L$ to the dynamical equation
\begin{eqnarray}\label{dynamical_equation}
\frac{dL_{\alpha\beta}}{dt} ~&=&~ \frac{d}{dt} R_{\alpha\rho} R_{\beta\sigma} I^S_{\rho, \sigma; \mu \nu} (\omega_{\mu \nu} + \Omega_{\mu \nu} ) \nonumber \\
~&=&~  R_{\alpha\rho} R_{\beta\sigma} \bigl(  \frac{d}{dt} I^S_{\rho, \sigma; \mu \nu} (\omega_{\mu \nu} + \Omega_{\mu \nu} ) \,  \nonumber \\
~&+&~ \, \omega_{\rho \kappa}  I^S_{\kappa, \sigma; \mu \nu} (\omega_{\mu \nu} + \Omega_{\mu \nu} )  \, + \, \omega_{\sigma \kappa}  I^S_{\rho, \kappa; \mu \nu} (\omega_{\mu \nu} + \Omega_{\mu \nu} ) \bigr) \nonumber \\
~&\equiv&~ R_{\alpha\rho} R_{\beta\sigma} \frac{D}{Dt}  I^S_{\rho, \sigma; \mu \nu} (\omega_{\mu \nu} + \Omega_{\mu \nu} )
\end{eqnarray}
The main virtue of bringing $R$ factors outside the time derivative on the right-hand side, as in the third line above, is that when the left-hand side vanishes we can simply strip them away.  

We can write Eqn.\,(\ref{dynamical_equation}) more compactly as
\begin{equation}
{\cal R}^{-1} \frac{dL}{dt} ~=~ \frac{D}{Dt} I^S (\omega + \Omega)
\end{equation}

 For some applications it might also be useful to go to a rotating frame, where 
 \begin{equation}
 x^j_{\rm rot.} (t) ~=~ V(t) x^j (t) 
 \end{equation}
 motivates use of the variable
 \begin{equation}
 R_{\rm rot.} (t) ~\equiv~ V(t) \,  R(t) \, V^{-1} (t) 
 \end{equation}
 so that 
 \begin{equation}
V \,  R \,  x^j ~=~ R_{\rm rot.} \, x^j_{\rm rot.}
\end{equation}
It is not difficult to translate the dynamical equations into this sort of frame, eliminating $R(t)$ in favor of $V(t)$ and $R_{\rm rot.}$.

\subsection{Three Dimensional Notation}

In three dimensions a special ``vector'' notation for the $A$ indices is possible (and standard).  It relies on use of the invariant $\epsilon_{a\mu \nu}$ symbol to eliminate antisymmetric pairs of indices in favor of single indices.  
Here I will simply record the definitions and the main results.  Let me remark that premature resort to this notation can tend to obscure the conceptual structure exposed above.  

With
\begin{equation}
I^S_{ab} ~\equiv~ \frac{1}{2} \epsilon_{a\alpha\beta} \epsilon_{b\rho\sigma} I^S_{\alpha\beta; \rho\sigma} ~=~ \sum\limits_j m^j \, (s^{j \, 2} \delta_{ab} - s^j_a s^j_b)
\end{equation}
\begin{eqnarray}
\omega_a ~&\equiv&~ \frac{1}{2} \epsilon_{a\alpha\beta} \omega_{\alpha\beta} \\
\Omega_a ~&\equiv&~ \frac{1}{2} \epsilon_{a\alpha\beta} \Omega_{\alpha\beta} 
\end{eqnarray}
we have
\begin{equation}
R^{-1}_{ab} L^b ~=~ I^S_{ac} (\omega_c + \Omega_c)
\end{equation}
and in an evident notation (suppressing indices)
\begin{eqnarray}
R^{-1} L ~&=&~ I^S (\omega + \Omega) \\
R^{-1} \frac{dL}{dt} ~&=&~ \frac{D}{Dt} I^S (\omega + \Omega) \label{3D_dynamical}
\end{eqnarray}
with
\begin{equation}
\frac{D}{Dt} ~=~ \frac{d}{dt} \, + \, \omega \times
\end{equation}

\subsection{Specializations}

Falling cats or divers can face the challenge of re-orienting their bodies in space despite starting from rest (zero angular momentum) and with no access to external torques.  They can address that challenge by changing their shape.  By doing so, they induce ${L}^S \neq 0$ and in consequence $\Omega \neq 0$.  From 
\begin{equation}
R^{-1} { L} ~=~ I^S (\omega + \Omega) ~=~ 0
\end{equation}
we deduce 
\begin{equation}\label{cat_equation}
\omega dt ~=~ - \Omega dt ~=~ - (I^S)^{-1} L^S dt 
\end{equation}
Here we see that motion in physical space reflects the gauge potential in shape space quite directly \cite{sw2}.  Indeed, the (ordered) time integral of angular velocity, which gives the referenced rotation in physical space, is equal in magnitude to the integral of the geometric potential, expressed as a one-form, in shape space.  

Several examples illustrating this physical phenomenon have been analyzed in detail \cite{al_student, alba, wisdom, batterman, antti, sw2, montgomery, littlejohn, cabrera}.  These examples establish the non-triviality of the shape space gauge structure, even for very simple families of shapes.   An interesting feature is the occurrence of topologically non-trivial gauge structures, including vortices and magnetic monopoles.  These can occur when the allowed shape space is topologically non-trivial.  Indeed, when they occur they indicate that the shape space cannot be augmented to become trivial without encountering configurations where $I^S$ is singular (that is, non-invertible).  It should be possible to analyze large classes of these singularities systematically.  

It could be interesting, also, to analyze cases where the shape space gauge structure plays a dominant but not exclusive role in the dynamics, with $L$ and $\frac{dL}{dt}$ introduced as small parameters or noise.   

At the other extreme, for a rigid body we can fix the standard shape once and for all.  In this case $L^S = 0$, and Eqn.~(\ref{3D_dynamical}) reduces to the classic Euler equations for rigid body motion.  

In some circumstances a specific choice of gauge suggests itself.  When we have a stable ordering of the eigenvalues $I_1, I_2, I_3$ of the inertia tensor, we can choose the eigen-directions to be oriented along the $\hat x, \hat y, \hat z$ directions for the corresponding reference shapes.  When the eigenvalues cross, of course, we will need to patch together complementary gauge choices.  Alternatively, if we are especially interested in small deformations of a specific shape, we can choose to make the gauge potential and its ``radial'' components along a choice of spokes emanating from that point vanish, in the style of Gaussian normal co-ordinates.  (See also below.)

\subsection{Angular Momentum and Energy}

For vanishing external torques, the first line of the dynamical equation Eqn.\,(\ref{dynamical_equation}) allows us to infer, from $L_{\alpha \beta} \frac{dL_{\alpha \beta}}{dt} = 0$,  the conservation law
\begin{equation}
(M^S_{\alpha\beta} + L^S_{\alpha\beta})(M^S_{\alpha\beta} + L^S_{\alpha\beta}) ~=~ (\omega + \Omega)_A I^S_{A;B} I^S_{B;C}  (\omega + \Omega)_C ~=~ {\rm constant}
\end{equation}
This is the covariant completion of the familiar conservation of $(I \omega)^2$ for untorqued rigid bodies.  

The absence of external torques does not imply energy conservation, however.  Indeed, in the envisaged applications to cats, divers, autonomous systems and micro-machines shape changes might be accomplished by exerting or dissipating work.  It is interesting, nevertheless, to consider how the kinetic energy can be expressed.

First let us appreciate the difficulty.  The kinetic energy associated with motion according to Eqn.\,(\ref{shape_to_space}) is
\begin{equation}
E_{\rm kin} ~=~ \frac{1}{2} \sum\limits_j \, m^j {\dot x}^j \cdot {\dot x}^j ~=~ \frac{1}{2} \sum\limits_j \,  m^j \bigl( \omega_{\alpha \beta} s^j_\beta \omega_{\alpha \gamma} s^j_\gamma \, + \, \omega_{\alpha \beta} (s^j_\alpha {\dot s}^j_\beta - s^j_\beta {\dot s}^j_\alpha) \, + \, {\dot s}^j \cdot {\dot s}^j \bigr)
\end{equation}
Here the first two terms on the right-hand side match the corresponding terms of the expansion of the covariant completion of the standard expression for rigid bodies
\begin{equation}
\frac{1}{2} I^S_{A;B} (\omega_A + \Omega_A)(\omega_B + \Omega_B) ~=~ \frac{1}{2}  I_{A;B} \omega_A \omega_B  \, + \, \omega_A L^S_A \, + \,  \frac{1}{2}  I_{A;B} \Omega_A \Omega_B 
\end{equation}
but the third terms are quite different.  

We can choose fix our gauge so that $\Omega = 0$ at a given shape.  {\it In that gauge\/}  we have a simple separation of the kinetic energy {\it at that shape\/} into spatial motion and shape space contributions:
\begin{eqnarray}
E_{\rm kin} ~&=&~ \frac{1}{2}  I_{A;B} \omega_A \omega_B  \, + \, E^S_{\rm kin} \nonumber \\
E^S_{\rm kin} ~&\equiv&~  \frac{1}{2} \sum\limits_j \, m^j {\dot s}^j \cdot {\dot s}^j
\end{eqnarray}
This separation cannot be achieved globally, however. Gauge curvature obstructs a clean separation between rotational and deformational energy. 

Since the discrepant terms are of second order in the rate of shape change, in the limit of slow shape changes we can neglect them.  The term that is linear in $\dot s$ gives a residual correction in the adiabatic limit.  Indeed, if we scale the rate of deformation by writing $t = T\tau$ inside the argument of $s(t)$, while holding its functional form fixed, and integrate between fixed shapes $s(a), s(b)$, then we have schematically
\begin{equation}
\int\limits^{s(b)}_{s(a)} \, dt {\dot s} (t) ~\sim~ \int\limits_{s(a)}^{s(b)} ds 
\end{equation}
but 
\begin{equation}
\int\limits_{s(a)}^{s(b)} \, dt \dot s (t)^2 ~\sim~ \int\limits_{s(a)}^{s(b)} \, T d{\tau} \frac{1}{T^2} (\frac{ds}{d\tau})^2 \, \propto \frac{1}{T} \, \rightarrow \, 0 \ {\rm as} \ T \rightarrow \infty
\end{equation}

\section{Extensions}

\subsection{Blobs, Media, and Swarms}

We have formulated our equations with reference to systems of particles, but of course the method carries over to continuous mass distributions as a limit.  Here it is significant that our effective description needs only the inertia tensor and the shape-space angular momentum (and, for energy, the shape-space energy) as input, and is otherwise independent of the details of deformation.  

The framework of reference shapes and gauge symmetry, associated above with rotations of a body in space, can be carried over to problems involving the displacement of a deformable body in a medium.  The appearance of gauge structure in connection with separation of infinitesimal angular velocity (and simple velocity) into spatial and deformation pieces is quite general, though its dynamical salience will vary from case to case.   It plays a central role in the description of self-propulsion at low Reynolds number \cite{sw3} and in biological locomotion \cite{goldman}. 

When one has many independently deformable bodies, each will have its own ``internal'' gauge potential.   The dynamics of a swarm will bring in interactions that couple the gauge structures, and in some circumstances might energetically favor small space-time gradations of shape.  Then we could have an emergent Yang-Mills description of the collective motion.

\subsection{Molecules and Nuclei}

Many features in the spectra of molecules and of nuclei can be interpreted in terms of motions that combine rotation and deformation \cite{spectroscopy}.   When the deformations can be considered as small vibrations about a definite shape, we can choose the gauge potential at that shape to vanish.  This is accomplished by introduction of ``Eckart co-ordinates'' \cite{eckart}, which separate the motions approximately.  Effects of curvature cannot be made to vanish altogether, however.   Informed use of gauge theory ideas should allow more systematic and accurate treatment of this and related problems.

\bigskip

{\it Acknowledgement}: I am happy to thank Antti Niemi and Al Shapere for stimulating discussions.  This work is supported by the U.S. Department of Energy under grant Contract  Number DE-SC0012567, by the European 
Research Council under grant 742104, and by the Swedish Research Council under Contract No. 335-2014-7424.

\newpage


{\bf Appendix: Direct Calculation}

\bigskip

We will now work Eqn.\,(\ref{torque_equation}) into its reduced form more straightforwardly, directly in three dimensions.  
We express the second time derivative $\frac{d^2}{dt^2} \, (R_{\gamma\sigma}s^j_\sigma)$ as a sum of three terms and process each of those terms into an appropriate form. 
We will often use the property
\begin{equation}\label{epsilon_property}
\epsilon_{\alpha \beta \gamma} R_{\beta\rho} R_{\gamma \sigma} ~=~ \epsilon_{\tau\rho\sigma}R^{-1}_{\tau\alpha} ~=~ \epsilon_{\tau\rho\sigma} R_{\alpha\tau}
\end{equation}
of rotations.  It expresses that they have unit determinant, and also expresses the vector character of the vector cross product.

The first term simplifies as
\begin{equation}\label{I_final}
{\rm I} ~=~ \epsilon_{\alpha\beta\gamma} \sum\limits_j \ m^j R_{\beta\rho}R_{\gamma\sigma} s^j_\rho{\ddot s}^j_\sigma   ~=~ R^{-1}_{\kappa\alpha} \,  {\dot L}^S_\kappa
\end{equation}
where
\begin{equation}
L^S_\kappa ~\equiv~  \epsilon_{\kappa\rho\sigma} \sum\limits_j \, m^j s^j_\rho \, {\dot s}^j_\sigma
\end{equation}
has the form of angular momentum in shape space. It seems appropriate to call it geometric angular momentum, since it relates to motion in shape space rather than directly in physical space.

The second term, similarly treated, becomes
\begin{eqnarray}
{\rm II} ~&=&~ 2 \, \epsilon_{\alpha \beta \gamma} R_{\beta \rho} {\dot R}_{\gamma\sigma} \, \sum\limits_j m^j s^j_\rho {\dot s}^j _\sigma \\ ~&=&~ 2\, \epsilon_{\kappa\rho\delta}R^{-1}_{\kappa\alpha} (R^{-1} \dot R)_{\delta \sigma} \, \sum\limits_j m^j s^j_\rho {\dot s}^j_\sigma
\end{eqnarray}

To simplify this further, we separate the parts that are symmetric and antisymmetric in $\rho, \sigma$.  For the symmetric part we have
\begin{eqnarray}
~&{}&~ \frac{1}{2} \sum\limits_j m^j (s^j_\rho {\dot s}^j_\sigma \, + \, {\dot s}^j_\rho s^j_\sigma) 
~=~ \frac{1}{2} \, \frac{d}{dt} \, (\sum\limits_j \, m^j \, s^j_\rho s^j_\sigma) \\
~&{}&~ \equiv \frac{1}{2} \,  \frac{d}{dt}{\tilde I^S}_{\rho \sigma}
\end{eqnarray} 
For the antisymmetric part we have
\begin{eqnarray}
~&{}&~\frac{1}{2} \sum\limits_j m^j (s^j_\rho {\dot s}^j_\sigma \, - \, {\dot s}^j_\rho s^j_\sigma) ~=~ \frac{1}{2} (\epsilon_{\mu\rho\sigma}\epsilon_{\mu \lambda\eta} ) \, \sum\limits_j \, m^j s^j_\lambda{\dot s}^j_\eta \\
~&{}&~  \equiv \frac{1}{2} \epsilon_{\mu\rho\sigma} \, L^S_\mu
\end{eqnarray}

Collecting terms, we have
\begin{equation}\label{II_draft}
{\rm II} ~=~ (\frac{d}{dt}{\tilde I}{}^S_{\rho\sigma} \, + \, \epsilon_{\mu\rho\sigma} L^S_\mu)  \, \epsilon_{\kappa\rho\delta} \,  R^{-1}_{\kappa\alpha} \, (R^{-1} \dot R)_{\delta\sigma} 
\end{equation}

To reduce the redundancy of $R^{-1} \dot R$, which is automatically antisymmetric, we introduce
\begin{equation}\label{R_omega}
(R^{-1} \dot R)_{\delta\sigma} ~\equiv~ - \epsilon_{\lambda\delta\sigma} \, \omega_\lambda ~=~ - (\dot {R^{-1}} R)_{\delta \sigma}
\end{equation}
(The conventional $-$ sign will prove convenient later, and the second equation here shows that it is related to whether we regard $R$ or $R^{-1}$ as primary.) Inserting this into Eqn.~(\ref{II_draft}), after some $\epsilon$-gymnastics we find
\begin{eqnarray}
{\rm II} ~&=&~ (\frac{d}{dt} I^S_{\rho\sigma} - \epsilon_{\mu\rho\sigma}L^S_\mu)\, R^{-1}_{\sigma \alpha} \omega_\rho \\
~&=&~  R^{-1}_{\sigma\alpha} \, \bigl( \frac{d}{dt} I^S_{\rho\sigma} \ \omega_\rho \, + \, (\omega\times L^S)_\sigma \bigr) \label{II_final}
\end{eqnarray}
where
\begin{equation}
I^S_{\alpha \beta} ~=~ \delta_{\alpha\beta} {\tilde I}^S_{\eta \eta} - {\tilde I}^S_{\alpha \beta}
\end{equation}
is the inertia tensor of the standard shape.

The third term, similarly treated, becomes
\begin{eqnarray}
{\rm III} ~&=&~ \epsilon_{\alpha\beta\gamma} \, R_{\beta \rho} \, {\ddot R}_{\gamma \sigma} \, {\tilde I}^S_{\rho\sigma} \\
~&=&~ \epsilon_{\kappa \rho \mu} \, R^{-1}_{\kappa \alpha} \, (R^{-1} \ddot R)_{\mu \sigma}  {\tilde I}^S_{\rho \sigma} \label{III_draft}
\end{eqnarray}
To reduce the redundancy now we use
\begin{equation}
R^{-1} {\ddot R} ~=~ (R^{-1}\dot R)^. + (R^{-1} \dot R)^2
\end{equation}
together with Eqn.~(\ref{R_omega}) and some  $\epsilon$-gymnastics including
\begin{equation}
(R^{-1} \dot R)^2_{\mu \sigma} ~=~ \epsilon_{\mu \tau \lambda} \epsilon_{\tau \sigma \eta} \omega_\lambda \omega_\eta ~=~ \omega_\mu \omega_\sigma - \omega^2 \delta_{\mu \sigma} 
\end{equation}
to arrive at 
\begin{equation}\label{III_final}
{\rm III} ~=~ R^{-1}_{\kappa \alpha} \, (I^S_{\lambda \kappa} \, \dot \omega_\lambda \, + \, \epsilon_{\kappa \mu \rho} \, \omega_\mu \, I^S_{\rho \sigma} \omega_\sigma )
\end{equation}

Assembling Eqns.~(\ref{I_final}, \ref{II_final}, \ref{III_final}) and multiplying by $R^{-1}$, we arrive at
\begin{equation}\label{final_draft}
R^{-1}_{\beta \alpha} \frac{dL_\alpha}{dt} ~=~ {\dot L}^S_\beta \, + \, (\omega \times L^S)_\beta \, + \, {\dot M}^S_\beta \, + \, (\omega \times M^S)_\beta
\end{equation}
where
\begin{equation}
M^S_\beta ~\equiv~ I^S_{\beta \sigma} \, \omega_\sigma
\end{equation}
is the angular momentum associated directly with motion in physical space.

Finally, by introducing the covariant derivative 
\begin{equation}
\frac{D}{Dt} ~\equiv~ \frac{d}{dt} \, + \, \omega \times
\end{equation}
and adopting standard vector notation, we arrive at the reduced equation
\begin{equation}\label{memorable}
R^{-1} \frac{d{\bf L}}{dt} ~=~ \frac{D {\bf (M^S + L^S)}}{Dt}
\end{equation}

\end{document}